\documentclass[longbibliography, prb,reprint,amsmath,amssymb,showpacs,superscriptaddress,floatfix]
{revtex4-1}
\usepackage[breaklinks=true,colorlinks,citecolor=blue,linkcolor=blue,urlcolor=blue]{hyperref}

\usepackage{epsfig,mathrsfs,color,latexsym,subfigure,marginnote}
\renewcommand{\BibitemShut}[1]{}

\begin{document}
\title{Electric field induced gap modification in ultrathin blue phosphorus}

\author{Barun Ghosh}
\affiliation{Department of Physics, Indian Institute of Technology Kanpur, Kanpur 208016, India}
\author{Suhas Nahas}
\affiliation{Dept. of Material Science and Engineering, Indian Institute of Technology Kanpur, Kanpur 208016, India}
\author{Somnath Bhowmick}\email[]{bsomnath@iitk.ac.in}
\affiliation{Dept. of Material Science and Engineering, Indian Institute of Technology Kanpur, Kanpur 208016, India}
\author{Amit Agarwal}
\email{amitag@iitk.ac.in}
\affiliation{Department of Physics, Indian Institute of Technology Kanpur, Kanpur 208016, India}

\date{\today}
\begin{abstract}
We investigate the possibility of band structure engineering in the recently predicted 2D layered form of blue phosphorus via an electric field (E$_z$) applied perpendicular to the layer(s). Using density functional theory, we study the effect of a transverse electric field in monolayer, as well as three differently stacked bilayer structures of blue phosphorus. We find that, for E$_z > 0.2$ V/\AA~ the direct energy gap at the $\Gamma$ point, which is much larger than the default indirect band gap of mono- and bilayer blue phosphorus, decreases linearly with the increasing electric field;  becomes comparable to the default indirect band gap at E$_z \approx 0.45~(0.35)$ V/\AA ~for monolayer (bilayers) and decreases further until the semiconductor to metal transition of 2D blue phosphorus takes place at E$_z\approx 0.7~ (0.5)$ V/\AA ~for monolayer (bilayers).  Calculated values of the electron and hole effective masses along various high symmetry directions in the reciprocal lattice suggests that the mobility of charge carriers is also influenced by the applied electric field.  
\end{abstract}
\maketitle

\section{Introduction}
Various 2D crystals such as graphene \cite{graphene1, RevModPhys.81.109, 9781139031080, Santos_Graphene}, silicene \cite{PhysRevB.50.14916, PhysRevB.76.075131, 1.3524215, silicene2, PhysRevLett.109.056804}, germanene \cite{PhysRevB.76.075131, germanane1,PhysRevLett.102.236804}, transition metal dichalcogenides (MoS$_2$, MoSe$_2$, WSe$_2$ etc.) \cite{TMD1,10.1038/nphys2942, Santos_MoS2}, are being actively explored for the post Si nanoelectronics era, for their promise of  aggressive channel length scalability on account of  reduced short channel effects \cite{review2D}. Field effect transistors (FET's) based on graphene \cite{graphene_tr}, and MoS$_2$ \cite{MoS2_tr}, have already been demonstrated with performance which is superior to conventional Si based FET's. Layered black phosphorus (dubbed phosphorene) has joined this exciting family of 2D crystals of late, and FET's based on few layer black phosphorus have been fabricated with I$_{\rm ON}$/I$_{OFF}$ current ratio of $\sim10^5$, and field effect mobility of $\sim 1000$ cm$^2$/Vs \cite{BP-transistor, phosphorene1, APL_Neto}. 

\begin{figure}
\begin{center}
\includegraphics[width=0.99 \linewidth]{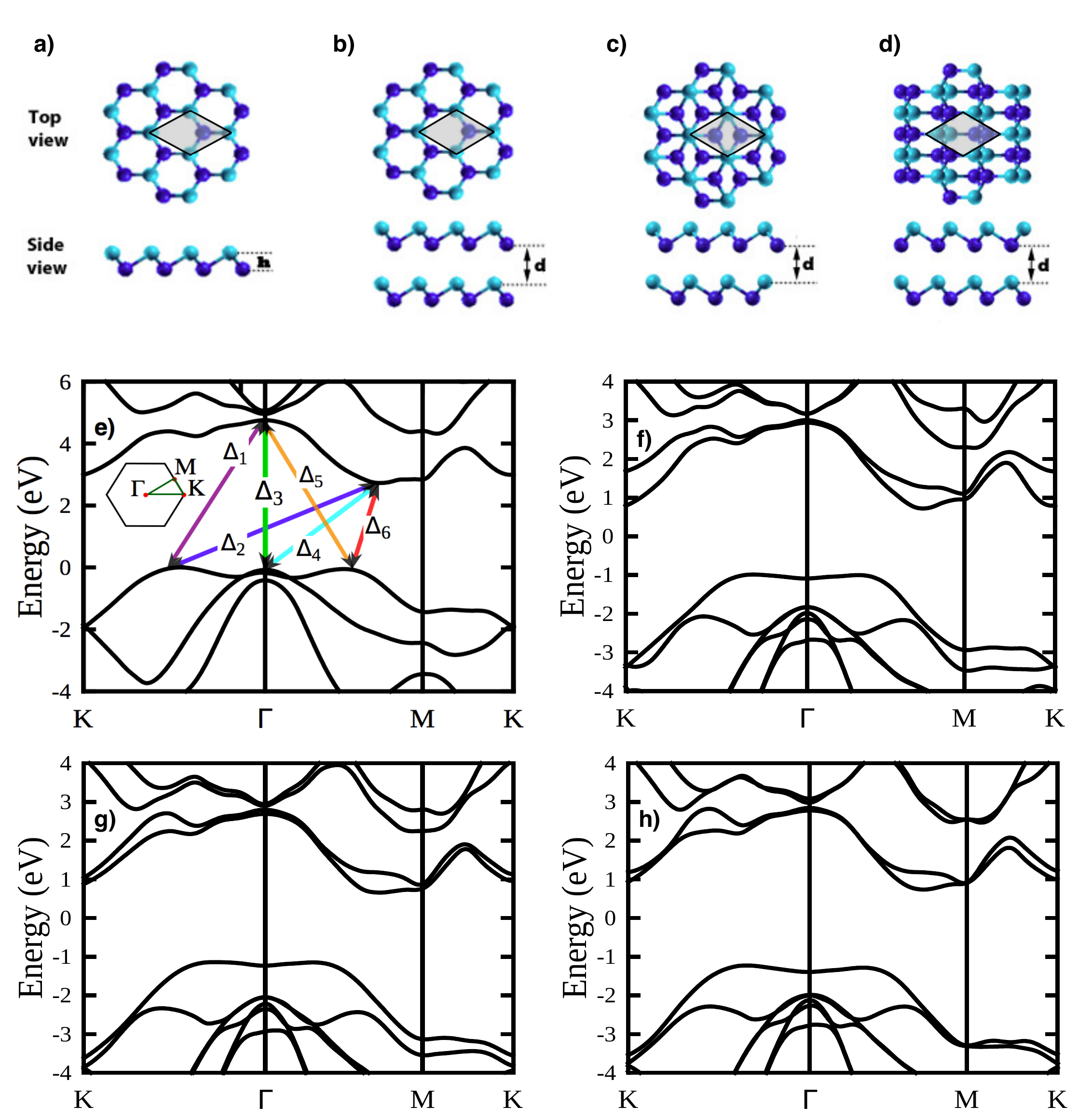}
\caption{ Panels (a), (b), (c) and (d) display the layered structure and panels (e), (f), (g) and (h) show the electronic band structure of monolayer, AA, AB and AC stacked bilayer blue phosphorus, respectively. Atoms located at the top and bottom of the quasiplanar layers are distinguished by cyan and dark blue color, respectively and they form two interpenetrating hexagonal sublattices. The shaded area in (a)-(d) represents the unit cell. Panel (e) defines different energy gaps between valence and conduction band, whose electric field dependence is illustrated in Fig.~\ref{f3}. The intrinsic indirect band gap is equal to $\triangle_2$ in all four structures. 
\label{f1}}
\end{center}
\end{figure}
 
More recently it has been predicted that, by certain dislocation of constituent P atoms, the puckered structure of black phosphorus (rectangular unit cell) can be converted to a more symmetric buckled structure (hexagonal unit cell) of another 2D allotrope, termed as the blue phosphorus \cite{zhu14, PRL.113.046804}. The newly proposed allotrope is actually a single layer of the A7 phase of phosphorus \cite{boul}. Based on first principles calculations, a single layer of blue phosphorus is predicted to be as stable as monolayer black phosphorus \cite{zhu14, PRL.113.046804}. Interestingly, electronic band structure of the two allotropes differ significantly. Unlike  monolayer black phosphorus, which has an intrinsic direct band gap of $\sim 1 $ eV at the $\Gamma$ point,   monolayer blue phosphorus has an indirect band gap of $\sim 2$ eV, based on GGA calculations which are well known to underestimate the band gap \cite{zhu14, PRL.113.046804}. For the case of phosphorene,  it has been shown that the band gap of black phosphorus depends on the number of layers \cite{phosphorene1}, strain \cite{PRLNeto} and applied electric field (perpendicular to the layers)\cite{blackp}. 
Such electronic band structure engineering in various 2D materials is essential for increasing their usefulness for application in various nano electronic and nano photonic devices. 
 
In this article we focus on electric field induced band gap modification in mono and bilayer blue phosphorus.  Such materials with  tunable band gap, particularly by means of a transverse electric field, would allow for flexibility and optimization of their usage in various devices. For example, in FET's  the gate voltage is applied perpendicular to the plane of the 2D material, which necessitate detailed understanding of the effect of transverse electric field on it's electronic band structure. The effect of transverse electric field on the electronic properties has earlier been studied both theoretically and experimentally for various 2D materials such as graphene \cite{PhysRevLett.99.216802, BiGexpt}, silicene \cite{silicene3, silicene1}, germanene \cite{silicene1}, transition metal dichalcogenides such as MoS$_2$ \cite{MoS2, PhysRevB.84.205325}, black phosphorus \cite{blackp} etc. 

For this article we use {\it ab initio} density functional theory calculations to study the electric field induced gap modification in mono and bilayer blue phosphorus with  three different stackings [see Fig.~\ref{f1}(a)-(d)]. All four (one monolayer and three bilayer) crystals, have an intrinsic indirect band gap. With  increasing electric field, the direct band gap at the $\Gamma$ point decreases almost linearly and becomes comparable to other indirect band gaps,  which on further increase of the electric field vanishes, leading to a insulator-metal transition.
The paper is organized in the following sequence: We describe the parameters used for DFT calculation and the crystal structure of monolayer and differently stacked bilayer blue phosphorus in Sec~\ref{sec2}. This is  followed by a detailed discussion of the dependence of the band gap and the electron and hole effective mass on the strength of transverse electric field in Sec~\ref{sec3}. Finally we summarize our results in Sec~\ref{sec4}. 

\begin{figure}[t]
\begin{center}
\includegraphics[width=.99 \linewidth]{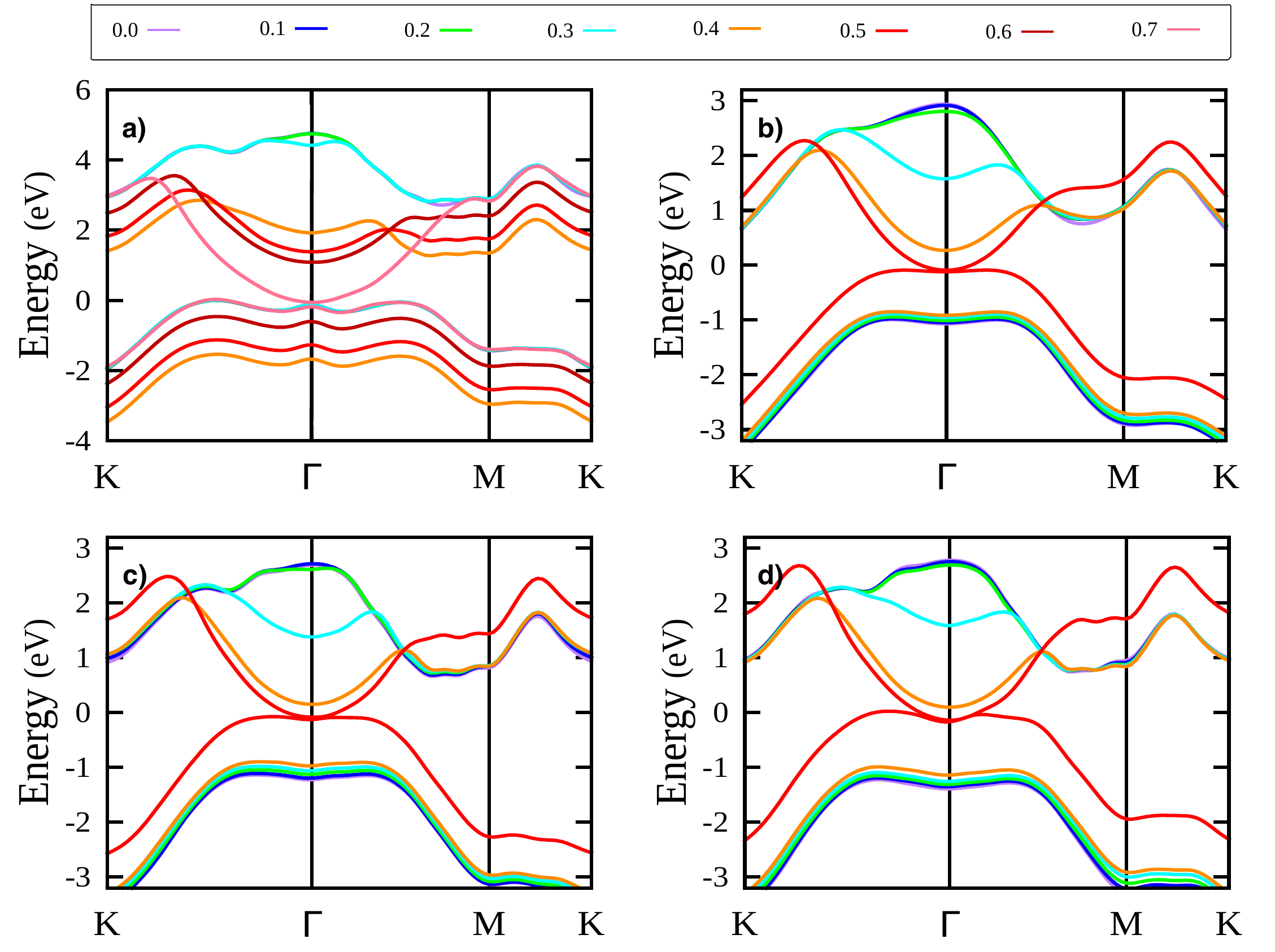}
\caption{  Panels (a), (b), (c) and (d) study the effect of the transverse electric field on the lowest conduction band and highest valance band for monolayer, AA, AB and AC stacked bilayer blue phosphorus, respectively. Note that for all four cases, the direct band gap at the $\Gamma$ point [$\triangle_3$, as defined in Fig.~\ref{f1}(e)] decreases with increasing electric field, while the default indirect band gap almost remains constant. The transition from widegap indirect semiconductor to a narrow gap direct semiconductor and eventually to a metal with increasing electric field is evident in all four plots. 	
\label{f2}}
\end{center}
\end{figure}   

\section{Crystal structure of monolayer and bilayer blue phosphorus}
\label{sec2}
The {\it ab initio} calculations are performed by density functional theory (DFT), using a plane-wave basis set and 
norm-conserving Troullier Martins pseudopotential, as implemented in Quantum Espresso \cite{QE}.  Electron exchange and correlation is treated within the framework of generalized gradient approximation (GGA-PBE) \cite{PBE} and a hybrid functional,  HSE06 \cite{HSE} . All the structures were optimized using the PBE functional coupled to DFT-D2 method as prescribed in Ref.~[\onlinecite{JCC:JCC20495}] to incorporate van der Waal's correction. 
Similar methods to relax black phosphorus have been shown to produce results that closely match experimental data \cite{Qiao}, thereby justifying our approach.  As a validation of our HSE06 calculations, we have reproduced the band structure of black phosphorene given in Ref.~[\onlinecite{Qiao}].
Since the trend and qualitative nature of all the results obtained using GGA and HSE06 are similar, and HSE06 has been demonstrated to give a better estimation of the electronic properties of black phosphorene \cite{Qiao}, only the HSE-06 based results are shown, unless stated otherwise. 
The kinetic energy cutoff for wave function is set to be 120 Ry.  A supercell with a vacuum of 25~\AA~ along the $z$ direction (perpendicular to the blue phosphorus layers) is constructed to eliminate the interaction with spurious replica images. A $k$-point grid of $12\times 12\times 1$ is used for the Brillouin zone integrations. Structural relaxations are carried out until the force on each atom (total energy change due to ionic relaxation between two successive steps) is less than $10^{-5}$ Ry/au ($10^{-6}$ Ry). An external sawtooth potential is used in the $z$ direction to simulate the effect of an applied electric field perpendicular to the blue phosphorus layer(s). 

We now describe the crystal structure of mono and bilayer blue phosphorus. The honeycomb net of P atoms in monolayer blue phosphorus is formed by two interpenetrating hexagonal sublattices [shown in two different colors/shades in Fig.~\ref{f1}(a)]. However, unlike graphene and similar to silicene/germanene, the two sublattices in blue phosphorus are not in the same plane and the vertical distance between them is given by $h$ [as shown Fig.~\ref{f1}(a), $h=$ 1.24~\AA]. This results in a pseudo-planar structure, where half of the atoms are buckled out of the plane, with a P-P bond angle equal to 92.9$^{\circ}$. Each of these monolayers can be stacked, so as to form multi layered structures in different ways, based on the relative displacement of the sublattices of one layer with respect to the other layers. In this paper, we consider bilayers stacked in three different sequences, namely AA [Fig.~\ref{f1}(b)], AB [Fig.~\ref{f1}(c)] and AC [Fig.~\ref{f1}(d)]. The unit cell has four atoms in total, two each in every layer. In case of AA stacking, the two sublattices belonging to each of the layers overlap directly and as a consequence, only two atoms per unit cell are visible in the top view [see Fig.~\ref{f1}(b)]. On the contrary, one constituent layer is displaced relative to the other one by a P-P bond length in case of AB stacking, such that one of the two sublattices of both the layers overlap directly and the other sublattice of the first layer is placed on top of the center of the honeycomb net of the second layer. Consequently, only three atoms per unit cell are visible in the top view of AB-stacked bilayer blue phosphorus, as shown in Fig.~\ref{f1}(c). Similar to AB, one layer of blue phosphorus is displaced with respect to the other one in case of AC stacking as well. However, unlike AB, none of the sublattices from the two layers are located directly on top of each other and thus, as illustrated in Fig.~\ref{f1}(d), all four atoms per unit cell are visible in the top view.  As reported in  Table~\ref{t1}, although the bond length and bond angle are nearly equal, irrespective of stacking sequence, the interplanar spacing $d$ changes significantly; for example AC (maximum $d$) has $6\%$ more interplanar spacing than AB (minimum $d$).
Similarly, total energy of bilayer blue phosphorus also depends on the stacking sequence, AA being energetically the most favorable one, followed by AC and AB, having 0.1 meV/atom and 2 meV/atom higher energy than AA.

\begin{table}[t]
\caption{Equilibrium structural parameters for monolayer and differently stacked bilayer blue phosphorus.}
\centering
\begin{tabular}{c c c c}
\hline
Stacking & Bond length & Bond angle & d \\
& (\AA) & (Degree) & (\AA)\\
\hline
Mono & 2.27 & 92.88 & -\\
AA & 2.26 & 93.06 & 3.23 \\
AB & 2.26 & 93.13 & 3.20 \\
AC & 2.27 & 93.00 & 3.41 \\
\hline
\end{tabular}
\label{t1}
\end{table}

Electronic band structure is calculated along various high symmetry directions in the reciprocal lattice [see the inset of Fig.~\ref{f1}(e)] and the energy dispersion relations are shown in Fig.~\ref{f1}(e)-(h). As illustrated in the figure, we define six different values of energy gap between the valence and conduction band, namely $\triangle_1$, $\triangle_2$, $\triangle_3$, $\triangle_4$, $\triangle_5$ and $\triangle_6$ and among them $\triangle_3$ is the only direct energy gap (measured at $\Gamma$ point). The choice is made on the basis of the fact that, the actual band gap of mono or bilayer blue phosphorus in zero or finite electric field is equal to one of the $\triangle_i$'s, as defined in Fig.~\ref{f1}(e). For example, in zero electric field single layer of blue phosphorus is an indirect band gap semiconductor and the band gap is equal to $\triangle_2$ [see Fig.~\ref{f1}(e)]. 
The electronic band structure is qualitatively similar for both GGA and HSE06 based calculations, although the magnitude of the band gap differs significantly; 1.94 eV (GGA) and 2.73 eV (HSE06), respectively  and the values for GGA compare well with those reported in the literature \cite{zhu14}. Similarly, bilayer blue phosphorus is also an indirect band gap semiconductor, and the band gap is equal to $\triangle_2$  for all three stacking sequences  (AA, AB and AC) although $\triangle_6\sim\triangle_2 \sim \triangle_4$ [see Fig.~\ref{f1}(f)-(h)]. Depending on the stacking sequence, the magnitude of the band gap of bilayer  blue phosphorus is  1.0 eV (GGA)  - 1.74 eV (HSE06) for AA , 0.97 eV (GGA) - 1.80 eV (HSE06) for AB, and 1.14 eV (GGA) -  1.97 eV (HSE06) for AC,  almost 40--50\% less than that of monolayer  blue phosphorus for all three bilayer structures. 

\begin{figure}[t]
\begin{center}
\includegraphics[width=.99 \linewidth]{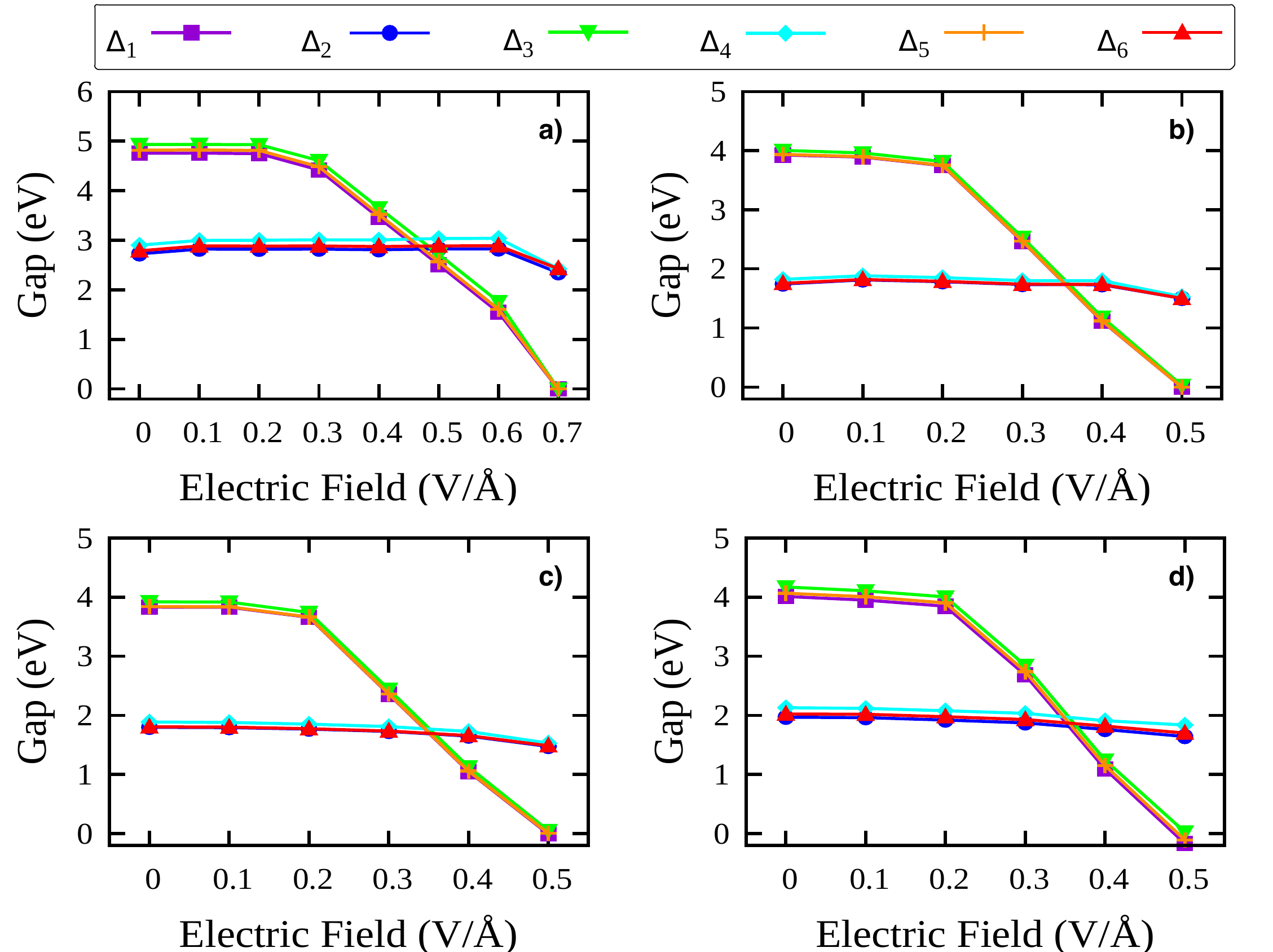}
\caption{ Panels (a), (b), (c) and (d) study the effect of the transverse electric field on the various band gaps specified in Fig.~\ref{f1}(e),  for monolayer, AA, AB and AC stacked bilayer blue phosphorus, respectively. Note that $\Delta_3$  (green triangle pointing downwards) denotes the direct band gap at the $\Gamma$ point, while all other $\Delta$'s represent indirect band-gaps. In all the panels, the direct band gap at the $\Gamma$ point decreases with increasing electric field, and becomes comparable to the indirect band gap, and both of which vanish for electric field strength of $\sim 0.7 ~(0.5)$ V/\AA, making monolayer (bilayer) blue phosphorus metallic.
\label{f3}}
\end{center}
\end{figure}

\section{Band gap and effective mass dependence on the transverse electric field}
\label{sec3}

We now explore the effect of an externally applied transverse electric field on the electronic band structure of monolayer and bilayer blue phosphorus.
For this purpose, an electric field is applied in a direction perpendicular to the blue phosphorus layer (ranging from 0.1 V/\AA~ to 0.7 V/\AA, in steps of 0.1 V/\AA), which is similar to applying a gate voltage in FET's. We find that, unlike silicene~\cite{silicene1}, electric field does not affect the buckling in the case of blue phosphorus (value of $h$ changes by 0.28\% when the strength of electric field is increased from 0 to 0.7 V/\AA) and other structural parameters like bond length and bond angle also do not change significantly. 
Although there is no considerable structural modification, electronic band structure changes substantially, particularly at higher value of applied electric field [see Figs.~\ref{f2}(a) and \ref{f3}(a)]. As shown in the figure, upto 0.2 V/\AA~ the electronic band structure of monolayer blue phosphorus is mostly unaffected by the applied field and the change is readily apparent only when the applied field is 0.3 V/\AA~ or more. Interestingly, the most significant change occurs at the conduction band edge at the $\Gamma$ point, which starts shifting towards the VBM and as a result the value of $\triangle_1$, $\triangle_3$ and $\triangle_5$ decreases with increasing electric field, while $\triangle_2$, $\triangle_4$ and $\triangle_6$ do not change appreciably [see Fig.~\ref{f3}(a)]. As illustrated in Fig.~\ref{f3}(a), monolayer blue phosphorus remains an indirect band gap (magnitude equal to $\triangle_2$) semiconductor upto 0.45 V/\AA~ applied field. However, when the electric field is increased further to 0.5 V/\AA, the downward shift of conduction band edge at the $\Gamma$ point is so large that the conduction band minima (CBM) is relocated to the $\Gamma$ point (from it's original position, in the middle of $\Gamma$-M), converting monolayer blue phosphorus to a semiconductor with comparable direct and indirect band gaps ($\triangle_3 \sim \triangle_1\sim \triangle_5$) [see Figs.~\ref{f2}(a) and \ref{f3}(a)]. Note that, the valence band maxima does not change significantly even at higher electric field; although it shifts to close the  band gap at 0.7 V/\AA~ applied field, valence band maxima (VBM) retains its location in the middle of the ${\rm K}-\Gamma $ direction.

\begin{figure}[t]
\begin{center}
\includegraphics[width=.99 \linewidth]{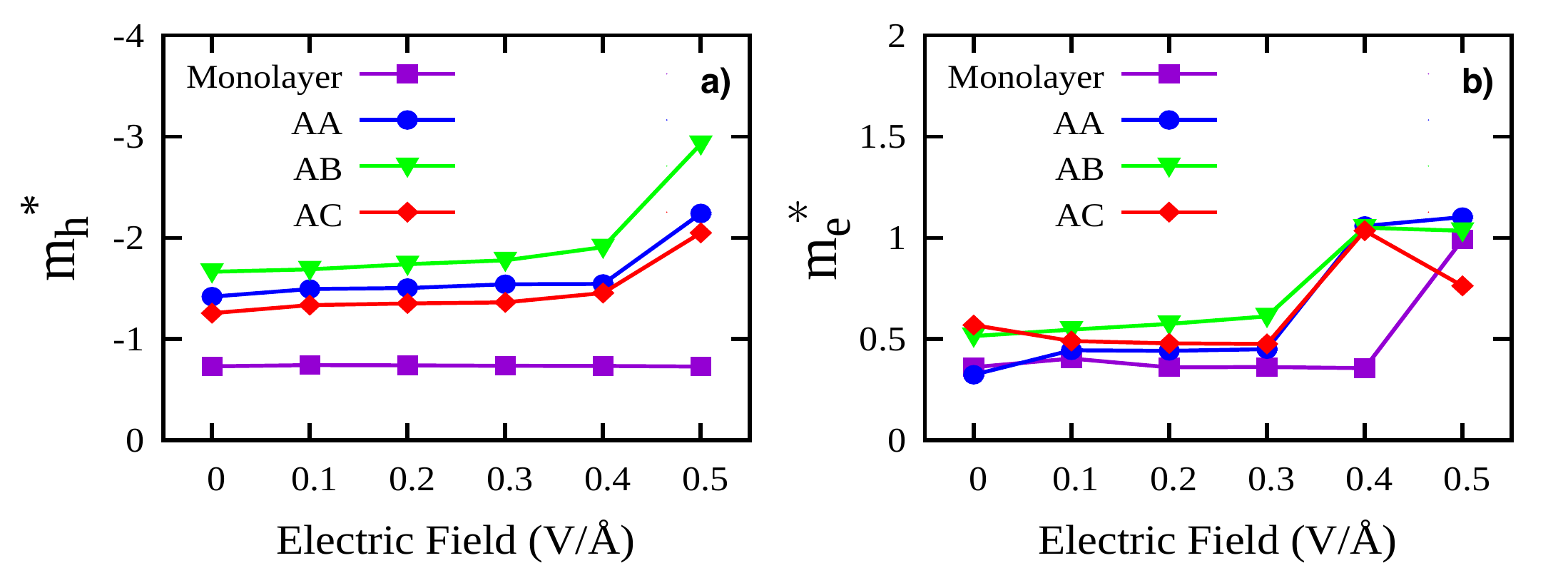}
\caption{ Panels (a) and (b) display the variation of hole and electron effective mass at valence band maxima (VBM) and conduction band minima (CBM) for mono layer and differently stacked bilayer blue phosphorus. If the VBM or CBM occurs in the middle of $\Gamma$M or $\Gamma$K, the effective mass is calculated in that particular direction. Otherwise, at the $\Gamma$ point, it is found to be isotropic.
\label{f4}}
\end{center}
\end{figure}

Similar to the monolayer, structural parameters of bilayer blue phosphorus, like bond length, bond angle and buckling ($h$) are hardly affected by externally applied electric field and interplanar spacing changes only by 3.22\% in AA, 4.5\% in AB and 2.35\% in AC, as the electric field increases from 0 to 0.5 V/\AA. 
Like a single layer of blue phosphorus, the most significant change to the electronic band structure also occurs at the conduction band edge at the $\Gamma$ point, which gradually shifts downward with increasing electric field, starting from 0.3 V/\AA~ and higher [see Figs.~\ref{f2}(b)-(d)]. As a result, the conduction band minima, which is originally located approximately in the middle of $\Gamma$-M, is shifted to the $\Gamma$ point at 0.35 V/\AA~ electric field [see Figs.~\ref{f2}(b)-(d)]. Consequently, the band gap changes from $\triangle_2$ to $\triangle_1$ (which is also comparable to $\triangle_3$ and  $\triangle_5$) in all three structures [see Figs.~\ref{f3}(b)-(d)].  

Having discussed the effect of transverse electric field on the band gap, we move on to the effective mass which determines the electron/hole mobility and thus also controls the transport properties. 
From the energy dispersion relation, the effective mass is calculated both at VBM ($m_h^*$) and CBM ($m_e^*$). As shown in Fig.~\ref{f4}(a), the effective mass of holes ($m_h^*$) does not change significantly as a function of the transverse electric field, since the shape of the valence bands is not affected by the transverse electric field [see Fig.~\ref{f2}]. However, depending on the number of layers and stacking sequence $m_e^*$ increases by a factor of 1.5 to 4 with increasing electric field [see Fig.~\ref{f4}(b)], which can be attributed to the shifting of CBM to the $\Gamma$ point from it's original location at zero or smaller electric field. 

\section{Conclusion}
\label{sec4}
 To summarize, we have explicitly shown that monolayer and bilayers of blue phosphorus, are versatile new 2D materials, in which the band gap can be tuned from $\sim 2.73$ eV (HSE06) [$\sim 1.74$ eV (HSE06)] for monolayer (bilayer)  to $0$ eV, by means of a transverse electric field. Broadly, we find that,  the intrinsic indirect band gap remains constant while the direct band gap at the $\Gamma$ point, and another two indirect band gaps decrease (almost linearly) with increasing electric field. Both monolayer and bi-layer blue phosphorus undergo a transition from indirect band gap insulator to a direct band gap insulator  for $E_z \ge 0.45~(0.35)$ V/\AA~ for monolayer (bilayers), and eventually to metal for $E_z \ge 0.7 ~(0.5) $ V/\AA ~for monolayer (bilayers). All these changes in the band structure are also reflected in the effective mass. 

\section*{Acknowledgements}  
We acknowledge funding from the DST INSPIRE Faculty Award,  DST Fasttrack Scheme for Young Scientist and the Faculty Initiation Grant by IIT Kanpur. We also thank Computer Center IITK for providing HPC facility. 

%

\end{document}